# Spectrum Sharing Dynamic Protection Area Neighborhoods for Radio Astronomy

Nicholas Papadopoulos
Computer Science
University of Colorado Boulder
Boulder, CO, USA
nicholas.papadopoulos@colorado.edu

Mark Lofquist
Computer Science
University of Colorado Boulder
Boulder, CO, USA
mark.lofquist@colorado.edu

Andrew W. Clegg
Spectrum Engineering
Google
Reston, VA, USA
aclegg@google.com

Kevin Gifford
Computer Science
University of Colorado Boulder
Boulder, CO, USA
kevin.gifford@colorado.edu

*Abstract*—To enforce incumbent protection through a spectrum access system (SAS) or future centralized shared spectrum system, dynamic protection area (DPA) neighborhood distances are employed. These distances are distance radii, in which citizen broadband radio service devices (CBSDs) are considered as potential interferers for the incumbent spectrum users. The goal of this paper is to create an algorithm to define DPA neighborhood distances for radio astronomy (RA) facilities with the intent to incorporate those distances into existing SASs and to adopt for future frameworks to increase national spectrum sharing. This paper first describes an algorithm to calculate sufficient neighborhood distances. Verifying this algorithm by recalculating previously calculated and currently used neighborhood distances for existing DPAs then proves its viability for extension to radio astronomy facilities. Applying the algorithm to the Hat Creek Radio Observatory (HCRO) with customized parameters results in distance recommendations, 112 kilometers for category A (devices with 30 dBm/10 MHz max EIRP) and 144 kilometers for category B (devices with 47 dBm/10MHz max EIRP), for HCRO's inclusion into a SAS and shows that the algorithm can be applied to RA facilities in general. Calculating these distances identifies currently used but likely out-of-date metrics and assumptions that should be revisited for the benefit of spectrum sharing.

## I. INTRODUCTION

Spectrum sharing is a vital part of the national economy by increasing "sustainable growth, innovation, and global competitiveness" [1]. One major system to manage spectrum is currently employed in the citizens broadband radio service (CBRS) band at 3,550 MHz to 3,700 MHz frequencies at the category A and B power levels (max 30 dBm/10 MHz and 47 dBm/10 MHz, respectively). CBRS employs a three-tiered hierarchy [2] system and a spectrum access system (SAS) [3], [4] to manage devices in the band. For spectrum access, devices are required to register with a SAS and request permission to transmit. The SAS can then grant permission on spectrum sharing parameters such as particular frequencies and time duration. Devices must also repeatedly request to continue transmitting, at which point the SAS may instruct the device to modify its transmission by instructing it to reduce transmission power, move to a different frequency band, or turn off completely.

In order to protect incumbents, which are usually federal licensees, in a particular dynamic protection area (DPA), the SAS must calculate the aggregate interference of devices that may interfere with the incumbent. Calculating this aggregate interference for all registered devices against all incumbents would be both intractable and unnecessary. Instead, the SAS considers only devices within some distance of the incumbent. This distance is called the neighborhood distance, where any CBSD within the neighborhood distance is considered in the calculation and is a candidate to be put on the "move list". CBSDs on the move list must either turn off, move outside of the neighborhood, or negotiate with the SAS to transmit at a lower power.

In Figure 1, the red line represents a DPA neighborhood distance. With the DPA point coordinates at the center of the pins, represented by the red circle, the white pins are within a 100 km neighborhood distance from the DPA antenna and are potential candidates for transmission alteration instructions. Devices up to 200 km and outside of the DPA neighborhood are shown in blue and may operate unhindered by SAS instruction.

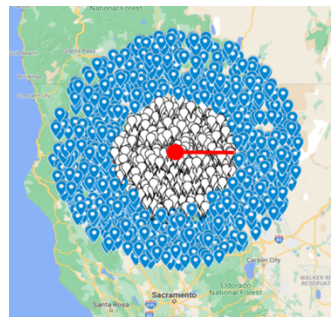

Fig. 1. Representation of a DPA neighborhood distance.

Before DPAs, CBSD deployments were based on exclusion zones. These zones were strict, and no devices were permitted to transmit within the bounds of the zone. Later, some of these zones were converted to DPAs, where the area would only be activated under certain circumstances. A DPA's activation means that it will enforce interference restrictions. If it is not activated, then devices are permitted to transmit without regard to interference to the incumbent, as a primary user. Some reasons to activate the area are the incumbent explicitly states that they will be using the area or environmental sensing capability (ESC) devices determine that the incumbent is using the area.

NTIA first calculated exclusion zones in 2015 using Monte Carlo analysis to predict the minimum distance where devices outside of the distance would not contribute interference above a threshold [5]. The exclusion zone was then defined by this distance, and devices outside of this zone would be very unlikely to cause a detectable amount of interference to the incumbent. The area within this distance was "excluded" as a transmittable area, and all devices within the zone were not allowed to transmit. These exclusion zone distances differ from current neighborhood distances, which consider devices that are within the zone as interferers, but those devices are not necessarily prevented from transmitting.

This strategy was very restrictive and oftentimes unnecessary. Even if the incumbent was not actively using the spectrum, other devices were not allowed to transmit in the zone. As a result, DPAs were developed in order to decrease exclusivity. In these areas, not all devices are excluded from using the spectrum, and none are even considered to be excluded if the incumbent is not using the spectrum. These DPAs were previously calculated by NTIA using modifications to the exclusion zone calculations. These results are embedded in mapping files (using keyhole markup language (KML))[1] used for current DPA calculations.

The novelty of this paper is to base an algorithm on these two methodologies with a goal of incorporating passive (receive-only) radio astronomy (RA) facilities into current SAS architecture progresses towards a further increase in spectrum sharing. Using the modified algorithm, as described in this paper, to recalculate current DPA neighborhood distances validates the algorithm and provides credibility to calculate DPA neighborhood distances for Hat Creek Radio Observatory (HCRO). The algorithm can then generalize neighborhood distance calculation for any RA facility to be incorporated into the SAS architecture in the future.

This process reviews and identifies current DPA neighborhood assumptions and metrics that may need to be reconsidered. Many of these metrics are overly conservative and can safely and realistically be adjusted to allow for smaller DPA neighborhoods and further spectrum sharing.

## II. Literature Review

"Spectrum sharing is a way to optimize the use of the airwaves, or wireless communications channels, by enabling multiple categories of users to safely share the same frequency bands," [6]. As airwaves are becoming more electromagnetically congested (and even contested) with the growing wireless communication industry, spectrum has become a valuable and scarce resource. Current methods to manage spectrum involve expensive auctions and exclusive use by a single party, even if the party is not using the spectrum. "Three-Tier Shared Spectrum, Shared Infrastructure, and a Path to 5G" by Preston Marshall [7] reviews current spectrum management practices along with current and near future solutions.

There are various architectures currently in place in an attempt to increase spectrum sharing. Among them, promising architectures are two- and three-tier architectures whose benefits have now been pragmatically demonstrated. A prominent three-tier architecture is deployed in the U.S.A. 's 3.55 GHz to 3.7 GHz frequency range, referred to as CBRS. This band is the first band in which a three-tier architecture has been deployed in the U.S.A. and which has been successfully managed by SASs since 2020 with no reported interference issues for incumbents.

CBRS uses a three-tier spectrum sharing framework that allows for shared use between incumbents, priority access licensed (PAL) users, and general authorized access (GAA) users. A SAS controls user access and may instruct users to move frequency bands or decrease their transmit power if the incumbent is using the same channel and unacceptable interference to the incumbent would occur. This is done by defining DPA neighborhoods, expressed as radii distances, around points where fixed incumbent users reside, inside of which devices may be considered to cause harmful interference to the incumbent. If the SAS determines that a transmitter is causing too much interference, it can provide power modification instructions to the transmitter. Devices that are outside of the DPA neighborhoods can continue their normal operation, as they are considered non-interferers.

Following the success of the CBRS SAS, the FCC has decided to allow a similar spectrum management system for the 6 GHz band, which refers to the frequency band between 5.925 GHz and 7.125 GHz. The band is currently, primarily inhabited by fixed services, which "provide microwave links for utilities, public safety, transportation and other [critical services]," [8] and will be opening to new users under the management system, termed an automated frequency controller (AFC). The 6 GHz band will, like a CBRS SAS, employ spectrum database coordination and will "coordinate at least outdoor deployments to insure no interference with tens of thousands of point-to-point microwave links and other incumbents" [9].

This paper centers on the creation of neighborhood distance for spectrum sharing in the CBRS band at HCRO and draws on a large body of research on the creation of the three-tier architecture in CBRS. Much of the work done for this project is based on the methodology outlined in NTIA Report 15-517 3.5 GHz Exclusion Zone Analyses and Methodology, which describes how exclusion zones were calculated for both ship and land-based radar sites [5]. The authors of this document outline the process for calculating exclusion zones around government radar sites, including how to use population assessments to estimate numbers of CBSD in a given area and how to factor in attenuation, among other considerations.

Since NTIA Report 15-517 is geared toward situations where government radar is the incumbent, the interference thresholds used in the report are not applicable to the radio astronomy sites considered in this paper, which have different operational metrics, such as a significant increase in sensitivity. To calculate a DPA neighborhood for a radio astronomy site, an interference threshold environmentally applicable to that site must first be determined. This threshold is based in part on the interference protection criteria (IPC) set for radio astronomy, defined by the ITU as the power that would introduce an error of 10% (-10 dB) into the smallest change in spectral power density that can be detected and measured by a receiver [10].

---

[1] https://www.ntia.doc.gov/fcc-filing/2015/ntia-letter-fcc-commercial-operations-3550-3650-mhz-band

In addition to NTIA Report 15-517, direct collaboration with team members at NTIA gave rise to numerous potential updates to the methodology and metrics used in the simulation. As defining DPAs is still a work in progress, NTIA is continually making updates, which may not always be publicized, so direct collaboration is crucial in keeping current methodology, assumptions, and metrics up-to-date.

### III. METHODOLOGY

#### A. Simulation Overview

The algorithm runs a Monte Carlo simulation to determine a neighborhood distance of a DPA (see Fig. 1). Each iteration deploys a number of CBSDs based on population density and distributes them with heights and coordinates within the given simulation radius from the DPA. The algorithm then calculates expected interference from each CBSD to the incumbent and employs a binary search for a minimum neighborhood distance. To do this calculation, the algorithm determines which CBSDs would be instructed to alter their transmission power for each tested neighborhood distance, i.e. put on the move list, by a standard SAS. The algorithm then reports the minimum distance in which the move list is the same as the maximum distance, which includes all of the simulated devices. The algorithm performs 1,000 iterations of this and reports the 95th percentile of the results as the recommended neighborhood distance.

There are two categories of devices for which a neighborhood distance must be calculated. Category A devices may not operate above 30 dBm, while category B devices may not operate above 47 dBm. When determining a category A distance, it assumes a maximum category B distance and varies the category A distance to test by multiples of 16 kilometers. When determining a category B distance, it assumes a maximum category A distance and performs the same search. Assuming maximum distances results in a bit of extra computation compared to using a realistic distance, but it eliminates the need to know the other category distance beforehand while maintaining accuracy, as the extra devices are considered non-contributors.

All parameters for the algorithm, such as AP heights and EIRP, were chosen based on assumptions for a rural region type. This reflects the region type of most, if not all, RA facilities will have.

Figure 2 represents the search resolutions for a neighborhood distance. Green dots represent category A APs while blue represent category B APs. The inner circle is the limit for which simulated category A APs were deployed, and the outer circle is the limit for which simulated category B APs were deployed. The distance between each red tick mark represents the 16 km search resolution. If searching for a category A neighborhood distance, all category B devices (blue dots) will be included in the aggregate interference regardless of distance. The category A neighborhood distance being tested will fall on one of the red tick marks within the inner circle, and the move list will be calculated while only considering category A devices within the distance being tested.

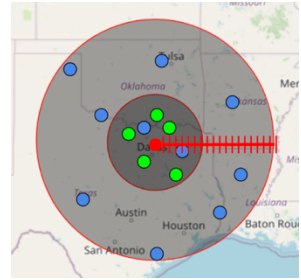

Fig. 2. Representation of the search resolutions for a neighborhood distance.

Figure 3 shows the effects of the tested neighborhood distance on the calculated move list and is an actual example of one iteration of the Monte Carlo simulation. The algorithm will choose the distance at which the slope of the line changes from positive to zero.

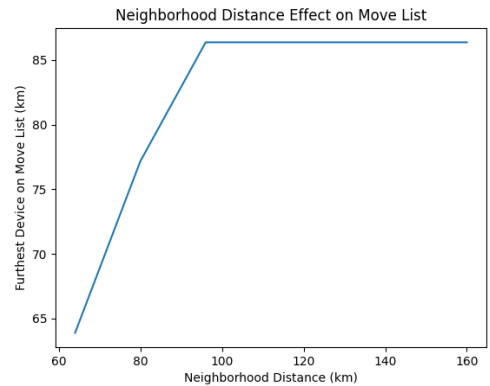

Fig. 3. Effect of the tested neighborhood distance on the calculated move list.

#### B. Number of CBSDs

The simulation first retrieves the population density of the simulation area of the DPA. The perimeter for the population density retrieval was a 200 kilometer radius for category A devices and a 500 kilometer radius for category B devices. The perimeter for each was determined using the Wireless Innovation Forum's method with a resolution of 100 arcseconds. This method retrieves the population density based on the 2010 census data[2].

The simulation calculates the number of UEs by multiplying the population density with the market penetration factor and channel scaling factor as described in [5], which are 0.2 and 0.1, respectively. An assumed 40% of these users are served by category A access points (AP), while 60% are served by category B APs. To determine the number of APs to simulate, the number of UEs for a category is divided by the number of UEs that are served by an AP of that category. The assumption is that three UEs are served by each category A AP and 500 UEs are served by each category B AP. Table I shows the numbers used for the current McKinney and the prospective HCRO DPAs. McKinney was chosen as a verification location because it is a current DPA that is most closely related to RA classifications, i.e. it is a single point DPA and is in a region classified as rural. These are two important classifications that can be expected for RA facilities. Single point DPA means that the DPA is defined by just a single pair of coordinates.

---

[2] https://www.sciencebase.gov/catalog/item/57753ebee4b07dd077c70868

TABLE I. NUMBER OF APs CALCULATION

|  | McKinney | | HCRO | |
|---|---|---|---|---|
| **CBSD Category** | **A** | **B** | **A** | **B** |
| Simulation Radius (km) | 200 | 500 | 200 | 500 |
| Population Density | 7,528,434 | 27,935,960 | 1,046,025 | 15,334,873 |
| Market Penetration | 0.2 | 0.2 | 0.2 | 0.2 |
| Channel Scaling | 0.1 | 0.1 | 0.1 | 0.1 |
| # of UEs | 150,569 | 558,719 | 20,921 | 306,697 |
| Users in CBSD Category | 40 | 60 | 40 | 60 |
| UEs in CBSD Category | 60,227 | 335,231 | 8,368 | 184,018 |
| UEs Served per AP | 3 | 500 | 3 | 500 |
| # of APs | 20,076 | 671 | 2,790 | 369 |

## C. CBSD Characteristics and Propagation Loss

*1) Propagation Loss:* The algorithm uses the Irregular Terrain Model (ITM) to calculate propagation loss for each CBSD. It uses the Wireless Innovation Forum parameters as inputs to the model, most notably a 0.1% - 99.9% reliability range. The model uses the 95th percentile of 2,000 Monte Carlo iterations with this reliability range as variance. The Wireless Innovation Forum ITM parameters are as follows:

TABLE II. ITM PARAMETERS

| *Polarization* | 1 (vertical) |
|---|---|
| *Dielectric constant* | 25 (good ground) |
| *Conductivity* | 0.02 S/m (good ground) |
| *Confidence* | 50% |
| *Reliability* | 0.1% - 99.9% |
| *Mode of Variability (MDVAR)* | 13 (broadcast point-to-point) |

The surface refractivity value varies by location and shall be derived by the methods and associated data files in ITU-R Recommendation P.452. The climate value varies by location and shall be derived by the methods and associated data files in ITU-R Recommendation P.617. Descriptions of most of the inputs can be found at [11] and [12].

*2) EIRP:* All category A APs were assumed to operate indoors, which includes a 15 dB building attenuation loss. Category A APs were all assumed to operate at 26 dBm/10MHz. All category B APs were assumed to operate outdoors at 47 dBm/10MHz. All APs in the simulation used a 6 dBi mainbeam antenna gain.

*3) Position:* Access points were each given coordinates within the CBSD category simulation radius with a uniformly random distribution of both bearing and distance from the DPA.

*4) Height:* The algorithm sets 80% of the category A APs at three meters tall and the other 20% at six meters tall. It sets category B APs at a uniform distribution between six and 100 meters tall.

## D. HCRO Interference Threshold Calculation

The following equation was used to determine the acceptable interference threshold. While the values can be modified as needed to any DPA facility, the following are the values for HCRO used in this project.

$$IT = TNF + NF_{Rx} + S + Loss_{Insertion} + I/N_{Ratio} - G_{Rx} \quad (1)$$

*1) TNF*: The thermal noise floor due to room temperature.

$$TNF = -174 \text{ dBm} \quad (2)$$

*2) $NF_{RX}$*: The noise figure of the receiver. This value can be found by taking the dB value of the value obtained from equation (3.20) in [13] using a temperature of 30 K [14].

$$NF_{Rx} = 10 \log_{10}(1 + \frac{T_{exc}}{T_0})$$
$$= 10 \log_{10}(1 + \frac{30}{290})$$
$$\approx 0.43 \text{ dB} \quad (3)$$

*3) S*: Scales the value in Hz to fill the channel bandwidth of 10 MHz.

$$S = 10 \log_{10}(10^7) = 70 \text{ dB} \quad (4)$$

*4) $Loss_{Insertion}$*: The insertion losses. This algorithm assumes a 2 dB insertion loss for both the receiver and the transmitter.

$$Loss_{Insertion} = 4 \text{ dB} \quad (5)$$

*5) $I/N_{Ratio}$*: The maximum acceptable interference to noise ratio (INR). This number was determined from [10].

$$I/N_{Ratio} = -10 \text{ dB} \quad (6)$$

6) $G_{Rx}$: The gain for the receive antenna. This is unique to any given antenna. Coordination with HCRO resulted in a 50 dB value for this parameter.

$$G_{Rx} = 50 \text{ dBi} \quad (7)$$

7) *IT*: The interference threshold for HCRO. Anything below this threshold can be considered undetectable by HCRO.

$$IT = -174 + 0.43 + 70 + 4 - 10 - 50 \approx -160 \text{ dBm} \quad (8)$$

### E. HCRO Antenna Characteristics

Characteristics for the HCRO antenna are found in [15] and presented in Table III.

TABLE III. HCRO ANTENNA CHARACTERISTICS

| | |
|---|---|
| *Location* | 40.81734, -121.46933 |
| *Height* | 6.1 meters |
| *Beamwidth* | 0.98° |
| *Azimuth range* | 0°-360° |

## IV. RESULTS

Table IV gives the neighborhood distances that the algorithm produced when running simulations for the current DPA at McKinney and the upcoming DPA at HCRO, while figures 4 and 5 show them graphically. One can see that the algorithm has recalculated the currently used distances for the McKinney DPA within two search resolutions of 16 kilometers. This is considered an accurate recalculation due to the variability in Monte Carlo analysis and the searching step size of 16 kilometers. This recalculation proves efficacy in the algorithm and in the calculated HCRO distances being reasonable. That is, if current SAS implementations were to include HCRO as a point DPA using a category A neighborhood distance of 112 kilometers and a category B neighborhood distance of 144 kilometers, one can expect that it will successfully be able to maintain CBSD interference below -160 dB. This should be undetectable by HCRO and will allow them to continue current operations with no noticeable effect.

By simply changing some parameters, this algorithm can apply to any RA facility in the U.S. The hope is to incorporate all RA facilities in the U.S. into SASs for the benefit of sharing national spectrum resources.

TABLE IV. CALCULATED NUMBER OF APS

| | *Calculated Category A (km)* | *Calculated Category B (km)* | *Current Category A (km)* | *Current Category B (km)* |
|---|---|---|---|---|
| McKinney | 160 | 448 | 150 | 416 |
| HCRO | 112 | 144 | N/A | N/A |

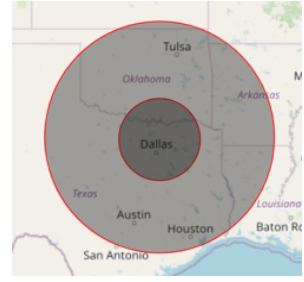

Fig. 4. The calculated category A and category B neighborhoods for McKinney, which are the inner and outer circles, respectively.

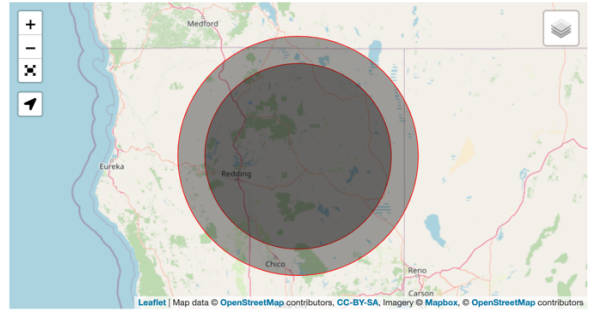

Fig. 5. The calculated category A and category B neighborhoods for Hat Creek Radio Observatory, which are the inner and outer circles, respectively.

## V. FUTURE WORK

Some of the assumptions and parameters in these calculations are either out of date or up for debate. Due to these potential inaccuracies, the current and reported DPA neighborhoods may be unnecessarily conservative, leading to overly complex computation and protective measures on CBSDs.

The interference to noise ratio is up for debate in the spectrum community. There is no clear evidence on which this criterion is based as no documentation was provided as to how this number was agreed upon. This -10 dB figure appears to have been determined somewhat arbitrarily [16]. While recent work NTIA's Technical Report TR-19-540 finds shortcomings with the current software simulations compared to measured harmful interference, further work on the simulation's receiver models could eventually improve measurements of IPC [17].

The market penetration factor in these calculations is now seen as out-of-date. Members of the community have noted that this is one metric that seems unnecessarily high when compared with the current data from SASs in use today.

Other metrics that can be gleaned from current SAS registration data is CBSD EIRP and heights. Rather than assuming maximum or near-maximum power from CBSDs and using an educated guess at their heights, more realistic data can be determined by actual grant information in use today.

The target for the interference threshold also has the potential to be drastically less conservative. Currently, this threshold is determined to prevent any detectable interference. However, one improvement could be made to determine a threshold that will still allow for readable reception as long as interference can be ignored and receiver components can operate without over saturation.

For convenience, Table V lists currently used metrics that are assumed or otherwise may be candidates for revision.

TABLE V. METRICS SUMMARY

| | |
|---:|:---|
| *Market penetration* | 0.2 |
| *Channel scaling* | 0.1 |
| *Users served by category A APs* | 40% |
| *Users served by category B APs* | 60% |
| *UEs per category A AP* | 3 |
| *UEs per category B AP* | 500 |
| *Category A EIRP* | 26 dBm/10 MHz |
| *Category B EIRP* | 47 dBm/10 MHz |
| *Building attenuation loss* | 15 dB |
| *Category A AP height* | 80% at 3 m; 20% at 6 m |
| *Category B AP height* | Uniformly random at 6 to 100 m |
| *Insertion losses* | 2 dB for both receiver / transmitter |
| *Interference to noise ratio* | -10 dB |
| *ITM Propagation* | Known to be overly conservative |

ACKNOWLEDGMENT

Nickolas LaSorte at NTIA was an integral part of building this algorithm. This work could not have been possible without his collaboration, work, and insight from previous efforts done at NTIA to calculate DPA neighborhoods.

Eloise Morris and Arvind Aradhya at University of Colorado Boulder along with Alexander Pollack at HCRO made essential contributions to this project. Their support in finding a reasonable threshold for HCRO made this project possible.

This material is based upon work supported by the National Science Foundation under Grant No. 2030233 and Grant No. 2139964. Any opinions, findings, and conclusions or recommendations expressed in this material are those of the author(s) and do not necessarily reflect the views of the National Science Foundation.

REFERENCES

[1] NITRD Wireless Spectrum R&D Senior Steering Group. "Promoting Economic Efficiency in Spectrum Use: the economic and policy research agenda." *NITRD*, 13 April 2013, https://www.nitrd.gov/pubs/WSRD_Workshop_IV_Report.pdf. Accessed 16 March 2022.

[2] The Bureau of Wireless Telecommunications. "3.5 GHz Band Overview." *Federal Communications Commission*, 10 March 2020, https://www.fcc.gov/35-ghz-band-overview. Accessed 16 March 2022.

[3] Spectrum Sharing Committee Work Group 3. "Signaling Protocols and Procedures for Citizens Broadband Radio Service (CBRS): Spectrum Access System (SAS) - Citizens Broadband Radio Service Device (CBSD) Interface Technical Specification." *Wireless Innovation Forum*, 25 November 2020, https://winnf.memberclicks.net/assets/CBRS/WINNF-TS-0016.pdf. Accessed 16 March 2022.

[4] Spectrum Sharing Committee Work Group 3. "Signaling Protocols and Procedures for Citizens Broadband Radio Service (CBRS): Spectrum Access System (SAS) - SAS Interface Technical Specification." *Wireless Innovation Forum*, 16 March 2020, https://winnf.memberclicks.net/assets/CBRS/WINNF-TS-0096.pdf. Accessed 16 March 2022.

[5] Drocella, E., et al. "3.5 GHz Exclusion Zone Analyses and Methodology." *National Telecommunications and Information Administration*, 18 June 2015, https://www.ntia.doc.gov/report/2015/35-ghz-exclusion-zone-analyses-and-methodology. Accessed 16 March 2022.

[6] "Spectrum Sharing." *NIST*, 4 February 2019, https://www.nist.gov/advanced-communications/spectrum-sharing. Accessed 21 April 2022.

[7] Marshall, Preston. *Three-Tier Shared Spectrum, Shared Infrastructure, and a Path to 5G*. Cambridge University Press, 2017.

[8] Alleven, Monica. "Federated Wireless gets ready to enable sharing in 6 GHz band." *Fierce Wireless*, 2020, https://www.fiercewireless.com/regulatory/federated-wireless-gets-ready-to-enable-sharing-6-ghz-band. Accessed 27 April 2022.

[9] Dynamic Spectrum Alliance, and Michael Calabrese. "Automated Frequency Coordination, An Established Tool for Modern Spectrum Management." *Dynamic Spectrum Alliance*, 12 March 2019, http://dynamicspectrumalliance.org/wp-content/uploads/2019/03/DSA_DB-Report_Final_03122019.pdf. Accessed 27 April 2022.

[10] "Protection Criteria Used for Radioastronomical Measurements." 2003. *ITU*, https://www.itu.int/dms_pubrec/itu-r/rec/ra/R-REC-RA.769-2-200305-I!!PDF-E.pdf. Accessed 11 12 2021.

[11] Henderson, Brian. "Radio Mobile - RF propagation simulation software - ITM model propagation settings." *Radio Mobile*, 2020, http://radiomobile.pe1mew.nl/?Calculations:ITM_model_propagation_settings. Accessed 21 March 2022.

[12] Tonkin, Susan. "A Tutorial on the Hata and ITM Propagation Models: Confidence, Reliability, and Clutter with Application to Interference Analysis." *SSRN*, 2018, https://papers.ssrn.com/sol3/papers.cfm?abstract_id=2880916. Accessed 21 3 2022.

[13] Rouphael, Tony J. *Wireless Receiver Architectures and Design: Antennas, RF, Synthesizers, Mixed Signal, and Digital Signal Processing*. Elsevier Science, 2014. *ScienceDirect*, https://www.sciencedirect.com/science/article/pii/B9780123786401000032. Accessed 21 March 2022.

[14] Welch, W.J., et al. "New Cooled Feeds for the Allen Telescope Array." *Publications of the Astronomical Society of the Pacific*, vol. 129, no. 045002, 2017. *IOPscience*, https://iopscience.iop.org/article/10.1088/1538-3873/aa5d4f/pdf. Accessed 21 March 2022.

[15] Harp, G. R., et al. "SETI OBSERVATIONS OF EXOPLANETS WITH THE ALLEN TELESCOPE ARRAY." *The Astronomical Journal*, vol. 152, no. 6, 2016, p. 181. *IOP Science*, https://iopscience.iop.org/article/10.3847/0004-6256/152/6/181. Accessed 28 March 2022.

[16] Federal Communications Commission. "Unlicensed Use of the 6 GHz Band." *The Federal Register*, vol. 85, no. 101, 2020, https://www.federalregister.gov/documents/2020/05/26/2020-11236/unlicensed-use-of-the-6-ghz-band.

[17] Achatz, R. J., and B. L. Bedford. "Interference Protection Criteria Simulation." *Institute for Telecommunication Sciences*, NTIA, Aug. 2019, https://www.its.bldrdoc.gov/publications/details.aspx?pub=3221. Accessed 21 April 2022.